\documentclass[12pt]{iopart}

\usepackage{graphicx}
\usepackage{color}

\definecolor{orange}{rgb}{.99,.6,.0}

\begin{document}
\title[Rare-gas clusters in intense VUV, XUV and soft x-ray pulses]{Rare-gas clusters in intense VUV, XUV and soft x-ray pulses: Signatures of the transition from nanoplasma-driven cluster expansion to Coulomb explosion in ion and electron spectra}

\author{Mathias Arbeiter and Thomas Fennel}

\address{Institute of Physics, University of Rostock, Universit\"atsplatz 3, 18051 Rostock, Germany}

\ead{thomas.fennel@uni-rostock.de}

\begin{abstract}
We investigate the wavelength dependent ionization, heating, and expansion dynamics of medium-sized rare-gas clusters (Ar$_{923}$) under intense femtosecond short-wavelength free electron laser pulses by quasi-classical molecular dynamics simulations. A comparison of the interaction dynamics for pulses with $\hbar\omega$=20, 38, and 90\,eV photon energy at fixed total excitation energy indicates a smooth transition from plasma-driven cluster expansion, where predominantly surface ions are expelled by hydrodynamic forces, to quasi-electrostatic behavior with almost pure Coulomb explosion. Corresponding signatures in the time-dependent cluster dynamics as well as in the final ion and electron spectra support that this transition is linked to a crossover in the electron emission processes. The resulting signatures in the electron spectra are shown to be even more reliable for identifying the cluster expansion mechanisms than ion energy spectra itself.
\end{abstract}

\section{Introduction}
Within the last decade, the rapid development of intense laser-matter science at short-wavelengths has been fueled by an enormous progress in free electron laser (FEL) technology~\cite{FEL,XFEL,SCSS,LCLS}. Today, the availability of high intensity pulses from the vacuum ultraviolet (VUV) over the extreme ultraviolet (XUV) up to the x-ray domain makes it possible to study nonlinear laser-matter processes in hitherto unexplored parameter regimes~\cite{Berrah2010}.
In particular the nonlinear response of nanosystems to intense FEL pulses has become a topic of great interest, as new fundamental insights into ultrafast laser-driven excitation and decay dynamics of many-particle systems can be gained~\cite{Saalmann2006,FennelRev2010}. Moreover, corresponding knowledge on the related radiation damage processes and timescales of target destruction is of central importance for novel applications, such as single-shot diffractive imaging of biological samples or time-resolved x-ray holography of highly excited nanosystems~\cite{Neutze2000,ChaN07,HauRiegePRL2010, NakamuraPRA2009}.

For studying nonlinear response processes of finite systems in intense short-wavelength laser radiation, atomic clusters have been proven to be versatile since the early days of the VUV-FEL at DESY~\cite{WabN02}. A key essence of the first VUV experiments and a considerable amount of successive theoretical work is, that collisional plasma heating through inverse Bremsstrahlung (IBS) is a major (typically the leading) heating process in the $\lambda\approx$100\,nm range and electron emission proceeds mostly via thermal evaporation, see e.g.~\cite{LaarmannPRL2005, SantraPRL2003, SiedschlagPRL2004, JungrJPB2005}. With increasing photon energy, however, collisional plasma heating processes diminish rapidly and different heating and electron emission processes take the lead.

\begin{figure}[h]
\begin{minipage}{1\columnwidth}
\begin{center}
\includegraphics[width=0.9\columnwidth]{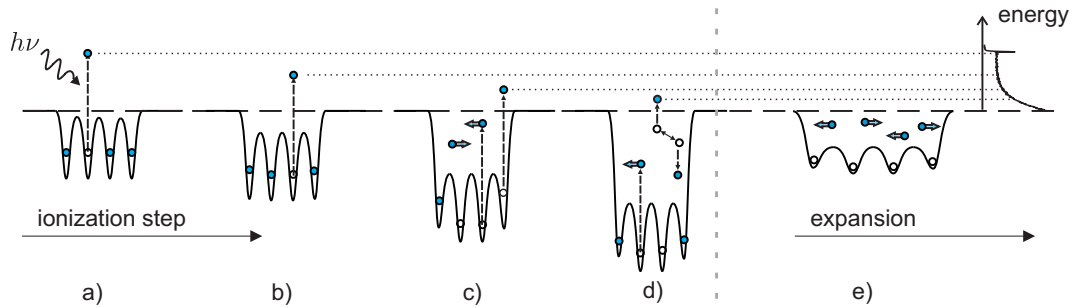}
\end{center}
\caption{\label{fig01}Schematics of the cluster ionization dynamics in intense short-wavelength laser pulses based on the effective cluster potential. After direct photoemission of the first electron (a), subsequently emitted photoelectrons experience a continuous Coulomb downshift with increasing cluster charge (b). This multistep ionization becomes frustrated at a certain ionization stage and nanoplasma formation sets in (c). Collisions between trapped electrons induce evaporation electron emission (d). Finally the cluster expands due to charging and hydrodynamic forces (e).}
\end{minipage}
\end{figure}

A key process at higher photon energy is sequential direct photoemission into the continuum~\cite{BosPRL08}. This so-called {\it multistep ionization} mechanism produces a characteristic plateau feature in the electron emission spectra and leaves behind an on-average homogenously charged cluster, as schematically sketched in Figs.~\ref{fig01}a-\ref{fig01}b.
Multistep ionization becomes suppressed, as soon as the cluster Coulomb potential produces an energy downshift larger than the atomic excess energy of the photoelectrons, see Fig.~\ref{fig01}c. This suppression effect begins at the cluster center (partial frustration) and then gradually spreads to the surface (full frustration). Simple estimates for the corresponding thresholds have been given in~\cite{ArbPRA2010}. If electron impact ionization can be neglected, which is assumed for our study, frustration of direct photoemission is required for efficient nanoplasma generation in the cluster (Fig.~\ref{fig01}d). Still, even if electron trapping sets in, collisional heating of the cluster-bound (quasifree) electrons remains negligible over a wide intensity range~\cite{Ziaja2009} and the thermal energy of the nanoplasma electrons is determined by inner ionization heating, i.e., by the excess energy from inner photoionization~\cite{ArbPRA2010}. From this stage on, energy equilibration through electron-electron collisions and evaporative electron emission yield an additional nearly thermal distribution to the multistep component of the electron spectrum~\cite{BosNJP2010, Gnodtke2010}, cf. Fig.~\ref{fig01}d. Hence, the cluster charging and the formation and energetics of the nanoplasma depend on the particular ionization regime and thus also determine the cluster expansion dynamics (Fig.~\ref{fig01}e)~\cite{Iwayama2009}. In the limit of pure multistep ionization, the quasi-homogenous charging of the system can be shown to induce a pure Coulomb explosion of the system. In contrast to that, ions in the inner region of the cluster can be efficiently screened in the presence of a dense nanoplasma, leading to a hydrodynamic expansion behavior~\cite{Murphy2008}. However, as we will show by examples, the identification of the expansion mechanism from ion energy spectra is often ambiguous. Alternative methods would thus be desirable for identifying the relative impact of the different explosion mechanisms.

The key motivation for the present study is the question, if electron spectra can be used for an easier identification of the leading expansion processes. Therefore we investigate the transition from hydrodynamic expansion to Coulomb explosion behavior as function of wavelength and examine the relations between the ionization regime and the electron and ion spectra. We analyze the cluster response for excitations with VUV, XUV and soft x-ray pulses using Ar$_{923}$ as a medium-sized model system. The applied molecular dynamics approach is similar to that of a previous study on the intensity-dependence of the electron emission from small rare-gas clusters at fixed wavelength~\cite{ArbPRA2010}.
For comparability of the simulations and to highlight the fundamental differences between the excitation regimes, the laser intensity is chosen such, that a fixed total energy absorption is achieved.
For extracting characteristic signatures from the ion spectra that can be used for comparison with experimental spectra, ion energy distributions are analyzed in two ways, i.e., shell-resolved as well as charge-state resolved. Therein, electron-ion recombination in the expanding cluster is taken into account, which is of particular interest for the investigation of core-shell effects~\cite{Thomas2009, Hoener2008, GnodtkePRA2009}.

The remaining text is structured as follows. The theoretical approach and its numerical implementation are described in Sec.\,\ref{sec:sim}. Simulation results are presented in Sec.\,\ref{sec:results}, containing an analysis from the time-dependent perspective (Sec.\,\ref{sec:results_td}) in relation to the final emission spectra (Sec.\,\ref{sec:results_spec}). Conclusions and an outlook are given in Sec.\,\ref{sec:summary}.

\section{Simulation method}
\label{sec:sim}
The laser-cluster interaction dynamics are modeled by a quasi-classical molecular dynamics approach. Therein, atomic ionization processes are described quantum mechanically via appropriate rates, whereas resulting ions and electrons is treated classically. This general strategy has been widely applied to laser-cluster interactions from the near infrared up to the x-ray domain, see~\cite{Saalmann2006,FennelRev2010} and references therein.

As optical field ionization is negligible in the wavelength range from the VUV to the soft x-ray range, single-photon ionization is considered to be the main inner ionization process. Departing from an initial Ar$_{923}$ cluster in relaxed icosahedral structure, photoionization of the cluster constituents is evaluated stochastically using single-photon ionization cross sections taken from free atomic Ar, see Tab.~\ref{tab01}.
\begin{table}[b]
\begin{center}
\begin{tabular}{l|cc|cc|cc}
\br
& occ$^{3s}$& occ$^{3p}$ & $I_p^{3s}$ [eV] & $\sigma^{3s}$ [Mb] (20/38/90 eV) & $I_p^{3p}$ [eV] & $\sigma^{3p}$ [Mb] (20/38/90 eV)\\
\mr
Ar&2&6&29.3& \begin{tabular}{p{0.4cm}p{0.1cm}p{0.4cm}p{0.1cm}p{0.4cm}} - & /& 0.2$^{[\rm a]}$ &/ & 0.2$^{[\rm a]}$ \end{tabular} & 15.76$^{[\rm b]}$ & \begin{tabular}{p{0.4cm}p{0.1cm}p{0.4cm}p{0.1cm}p{0.4cm}} 30$^{[\rm a]}$ & /& 5.0$^{[\rm a]}$ &/ & 1.2$^{[\rm a]}$ \end{tabular}\\

Ar$^{+}$ & 2 & 5 & 43.67 & \begin{tabular}{p{0.4cm}p{0.1cm}p{0.4cm}p{0.1cm}p{0.4cm}} - & /& - &/ & 0.2 \end{tabular} & 27.63$^{[b]}$ & \begin{tabular}{p{0.4cm}p{0.1cm}p{0.4cm}p{0.1cm}p{0.4cm}} - & /& 2.6$^{[\rm c]}$ &/ &  1.0 \end{tabular}\\

Ar$^{2+}$ & 2 & 4 & 58.09 & \begin{tabular}{p{0.4cm}p{0.1cm}p{0.4cm}p{0.1cm}p{0.4cm}} - & /& - &/ & 0.2 \end{tabular} & 42.54 & \begin{tabular}{p{0.4cm}p{0.1cm}p{0.4cm}p{0.1cm}p{0.4cm}} - & /& - &/ & 0.8 \end{tabular}\\

Ar$^{3+}$ & 2 & 3 & 73.60 & \begin{tabular}{p{0.4cm}p{0.1cm}p{0.4cm}p{0.1cm}p{0.4cm}} - & /& - &/ & 0.2 \end{tabular} & 57.60 &\begin{tabular}{p{0.4cm}p{0.1cm}p{0.4cm}p{0.1cm}p{0.4cm}} - & /& - &/ & 0.6 \end{tabular}\\

Ar$^{4+}$ & 2 & 2 & 90.07 & \begin{tabular}{p{0.4cm}p{0.1cm}p{0.4cm}p{0.1cm}p{0.4cm}} - & /& - &/ & - \end{tabular} & 74.96 &\begin{tabular}{p{0.4cm}p{0.1cm}p{0.4cm}p{0.1cm}p{0.4cm}} - & /& - &/ & 0.4 \end{tabular}\\

Ar$^{5+}$ & 2 & 1 & 107.4 & \begin{tabular}{p{0.4cm}p{0.1cm}p{0.4cm}p{0.1cm}p{0.4cm}} - & /& - &/ & - \end{tabular} & 90.94 & \begin{tabular}{p{0.4cm}p{0.1cm}p{0.4cm}p{0.1cm}p{0.4cm}} - & /& - &/ & - \end{tabular}\\
\br
\end{tabular}
\end{center}
\caption{\label{tab01}Photoionization cross-sections ($\sigma^{\alpha}$) and ionization energies ($I_p^{\alpha}$) for atomic Ar. Superscripts $\alpha$ indicates electron removal from 3s and 3p shell (as indicated) with initial shell occupation ${\rm occ}^{\alpha}$. Remaining ionization potentials are calculated with an atomic all-electron Dirac-LDA code~\cite{Ank1996}. Remaining cross sections have been extrapolated from the next known values assuming linear scaling with shell occupation.\\
$^{[\rm a]}$ Values taken from Ref.~\cite{BeckerShirley} \\
$^{[\rm b]}$ Values taken from NIST\\
$^{[\rm c]}$ Values taken from Ref.~\cite{Covington2001}\\
}
\end{table}
It should be noted that medium-induced atomic ionization threshold lowering due to screening and local plasma field effects is explicitly taken into account~\cite{FenPRL07b} and enables charging of cluster constituents beyond the maximum charge state of the atomic species, for details see~\cite{SiedschlagPRL2004, BauerAPL2004, GeoPRA07}.
The photoionization probability is determined from the instantaneous laser intensity $I(t) = I_0 f^2(t)$, where $I_0$ is the peak intensity and $f(t) = \exp(-2 \ln2 \: t^2 /\tau^2)$ is the normalized temporal field envelope of a Gaussian pulse with a pulse duration $\tau$ (FWHM). Resulting ions and electrons are propagated classically in the linearly polarized laser field (dipole approximation) and under the influence of a regularized Coulomb interaction of the form
\begin{equation}
V_{ij}(r_{ij},q_1,q_2)=\frac{e^2}{4\pi \varepsilon_0}\frac{q_i q_j}{r_{ij}} \,{\rm erf}\left(\frac{r_{ij}}{s}\right),
\end{equation}
where $e$ is the elementary charge, $r_{ij}$ is the distance between the interacting particles with charge state $q_i$ and $q_j$ and $s=1.128\,\AA$ is a numerical smoothing parameter. The latter prevents classical electron-ion recombination below the lowest quantum energy level. The classical trajectories are integrated using a standard velocity-Verlet algorithm. For the classification of active electrons we define a single particle energy
\begin{equation}
 E_i^{\rm sp}=\frac{m_i}{2}v_i^2+\sum_{i\ne j}V_{ij},
\end{equation}
where $m_i$ and $v_i$ are the mass and velocity of the $i$-th particle.
Electrons are denoted as quasifree (delocalized, still bound to the cluster) if $E_{\rm sp}<0$ and as free if $E_{\rm sp}\ge 0$ (continuum energy).

In order to calculate final ion charge states, e.g. to analyze charge-state dependent energy spectra, recombination of quasifree cluster electrons with ions is treated by the scheme developed in Ref.~\cite{FenPRL07b}. As radiative recombination rates are negligibly small for rapidly expanding clusters, only three-body-recombination (TBR) is resolved in our model. Since TBR proceeds mainly to highly excited, Rydberg-like atomic levels, a classical description is justified. To evaluate TBR, electrons are treated as recombined when localized to an ionic cell after a sufficiently long propagation time (here we use $t_{\rm recomb}=3$\,ps). The charge state of the corresponding ion is reduced by the number of  localized electrons. Remaining quasifree electrons persist and are assumed to be removed in an experiment by the extraction fields of the ion detector.

\section{Results and discussion}
\label{sec:results}
For examining the wavelength-dependent cluster response we compare excitations of Ar$_{923}$ with $\tau$=30\,fs laser pulses at \mbox{$\hbar \omega=20$}, 38, and 90\,eV, representing typical parameters presently available at FEL light sources. For brevity these cases will henceforward be denoted as VUV, XUV and soft x-ray. To achieve a fixed total energy absorption (13\,keV), the pulse intensities are chosen as \mbox{$I_0=2.5\times 10^{12}$}, $1.5\times 10^{13}$, and $5\times 10^{13}{\rm W/cm^2}$, respectively. The intensity increase with photon energy reflects the reduced photoionization cross sections at shorter wavelength.

\subsection{Time-dependent analysis}
\label{sec:results_td}
\begin{figure}[h]
\begin{minipage}{1\columnwidth}
\begin{center}
\includegraphics[width=1\columnwidth]{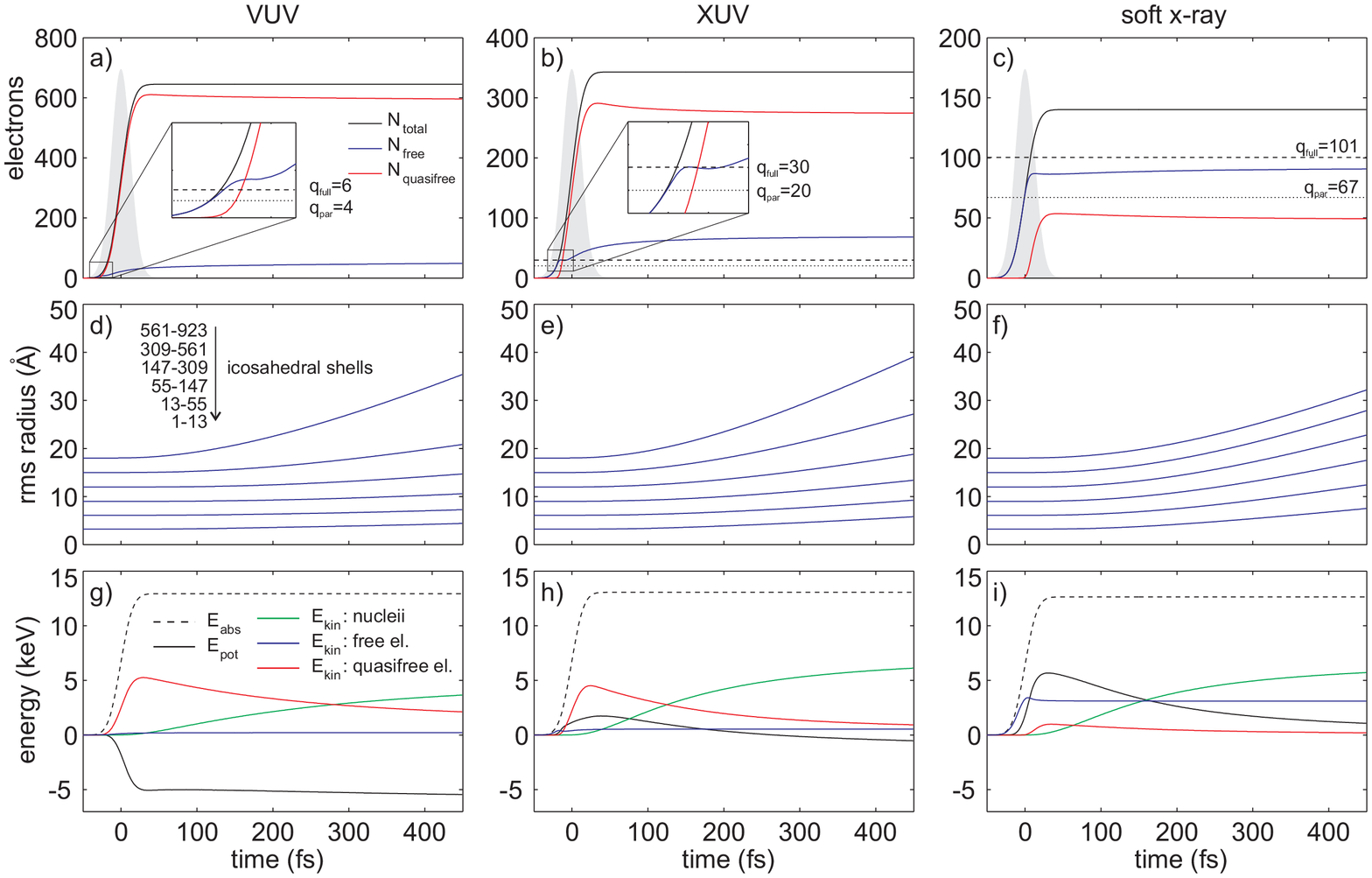}
\end{center}
\caption{\label{fig02}Calculated dynamics of Ar$_{923}$ under 30\,fs laser pulses at $\hbar \omega=20\,{\rm eV}$ (left), 38\,eV (middle), and 90\,eV (right) with laser intensities of \mbox{$I_0=2.5\times 10^{12}$}, $1.5\times 10^{13}$, and $5\times 10^{13}{\rm W/cm^2}$, respectively; (a)-(c): evolutions of the number of free and quasifree electrons (as indicated) together with corresponding frustration levels (dashed and dotted lines, see text); shaded areas indicate laser intensity envelopes; (d)-(f): shell-resolved cluster expansion based on the rms radii of the radial shells (as indicated); (g)-(i): time-evolution of absorbed, potential and kinetic energies (as indicated). }
\end{minipage}
\end{figure}

A time-dependent analysis of characteristic quantities as extracted from the three simulation scenarios is presented in Fig.~\ref{fig02}. For sufficient statistics we performed ensemble averaging over a sufficient number of runs. Thus all observables reflect statistical mean values. For all cases, the estimated critical cluster charge for partial ($q_{\rm par}$) and full frustration ($q_{\rm full}$) of direct photoionization due to the cluster Coulomb field are indicated assuming a spherical homogenously charged cluster with radius $R_{\rm cl}=19.6\,{\rm \AA}$ (radius determined from the root-mean-square radius of the ions) and 3p photoemission from neutral atomic Ar with an ionization potential of $I_p=15.76$\,eV. The number of free electrons in the simulations is taken as a measure for the outer cluster charge state.

The evolutions of cluster ionization displayed in Figs.~\ref{fig02}a-\ref{fig02}c show an increase of the final number of free electrons (blue curves) with photon energy, though the total number of activated electrons decreases (black curves).
Focussing on the VUV case ($\hbar\omega=20\,$eV), direct photoionization is operational only in a very early period of the laser pulse before the critical charge state for full frustration of $q\approx6$ is reached, see Fig.~\ref{fig02}a. As inner ionization continues, a dense nanoplasma is formed (red curve). Therein, collisional equilibration of quasifree electrons leads to thermal electron evaporation, which is the dominant emission process in this case. However, the vast majority of activated electrons remains trapped in the cluster potential. The electron spill-out produced by hydrodynamic electron pressure and the concentration of positive charge at the cluster surface result in a cluster potential that expels ions predominantly from the outer shells, see Fig.~\ref{fig02}d. Ions in inner shells hardly move within the displayed time interval, indicating efficient screening of the corresponding ions by the remaining quasifree electrons.

The ionization and expansion behavior gradually changes with increasing photon energy. In the intermediate XUV case ($\hbar\omega=38$\,eV), a notable period of multistep ionization occurs in the leading edge of the pulse, where only free electrons are created (Fig.~\ref{fig02}b). This direct emission stops abruptly as soon as the full frustration threshold is reached at about \mbox{$t\approx-15\,{\rm fs}$}, see Fig.~\ref{fig02}b. The subsequent generation of quasifree electrons again induces evaporative electron emission, which roughly sets in at time zero and produces a number of free electrons similar to the previous multistep ionization phase. In contrast to the ion dynamics in the VUV case, a much more pronounced expansion is observed for the second outer shell and also the inner shells show a considerable expansion effect, see Fig.~\ref{fig02}e.

In the soft x-ray case ($\hbar\omega=90$\,eV), multistep ionization is even more pronounced and frustration of direct emission begins only after the pulse peak, see Fig.~\ref{fig02}b. The subsequently produced quasifree electrons are less abundant than free electrons and only very weak evaporative emission is observed. It should be noted that the direct photoemission stops on a level that is slightly below the predicted threshold for full frustration (dashed line). This effect is attributed to the additional trapping field resulting from quasifree electrons that are produce in the inner region of the cluster as soon as the partial frustration level is reached. Because of their high excess energy (about 74\,eV for 3$p$ electron detachment from Ar), these electrons  induce a significant charge separation at the surface (spill-out) that enhances electron trapping. Further, all ionic shells expand similar to a pure electrostatic Coulomb explosion of  the cluster (Fig.~\ref{fig02}f).

Deeper insights into the importance of the different expansion mechanisms, i.e., the contributions from hydrodynamic vs. electrostatic Coulomb forces to the final total ion kinetic energy, can be extracted from a time-resolved analysis of the energy distribution in the system, see Figs.~\ref{fig02}g-\ref{fig02}i. The evolution of the total potential Coulomb energy and the kinetic energy contributions from ions and active electrons can be used to reconstruct the energy exchange processes. After the laser pulse. i.e., without further inner ionization, the sum of these components is constant because of energy conservation (not shown). It is further assumed, that the change of the kinetic energy of free electrons is negligible, which is fulfilled in good approximation in all scenarios discussed here. Under these circumstances, any gain of ion kinetic energy is either due to the release of thermal energy by expansion cooling of quasifree electrons (hydrodynamic) or due to the release of potential Coulomb energy (Coulomb explosion).

Based on these assumptions, the following picture can be deduced from the above scenarios. For the VUV case, the ion energy gain results almost completely from the release of thermal energy of quasifree electrons, indicating the hydrodynamic expansion regime, cf. Fig.~\ref{fig02}g. The small decrease of potential energy indicates that electrostatic Coulomb explosion is only a minor effect. As IBS-heating is negligible for all runs presented here (well below 1\%), the available thermal energy of quasifree electrons is determined by ionization heating. Note that significant IBS heating may occur for VUV excitation, but only at much higher pulse fluence~\cite{GeoPRA07}.

An intermediate situation is found with XUV excitation, cf. Fig.~\ref{fig02}h, where nearly equal release of thermal electron energy and potential energy occurs. This scenario thus represents a dynamical mixture of hydrodynamic expansion and Coulomb explosion with comparable contributions. Finally, nearly pure Coulomb explosion is observed in the soft x-ray case (Fig.~\ref{fig02}i), where potential energy release yields the main contribution to the ion recoil energy. The three different scenarios thus reflect the transition from hydrodynamic to Coulomb-driven cluster expansion.

\subsection{Electron and ion spectra}
\label{sec:results_spec}
In the next step, the final electron and ion spectra are examined in detail to extract characteristic features of the different ionization and expansion regimes. The spectra are sampled after 3\,ps of propagation and include averaging over a large set of simulation runs to provide sufficient statistics.

\begin{figure}[t]
\begin{minipage}{1\columnwidth}
\begin{center}
\includegraphics[width=1\columnwidth]{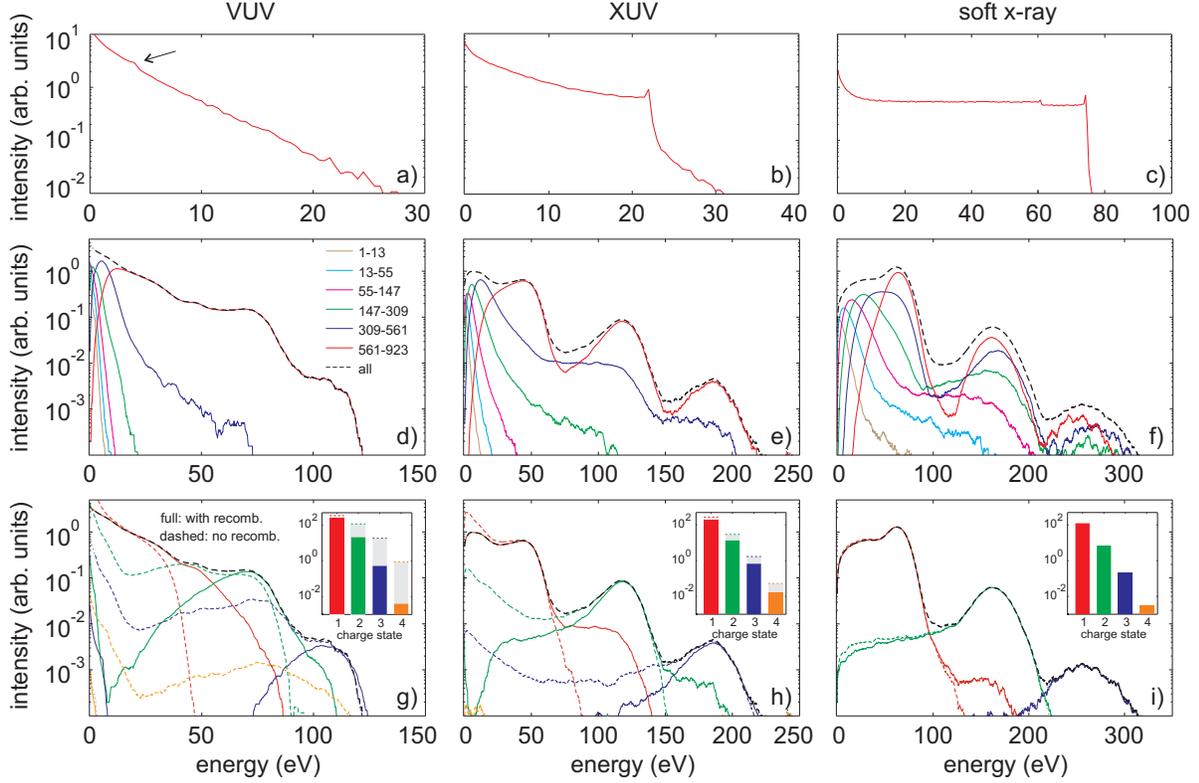}
\end{center}
\caption{\label{fig03}Calculated electron and ion spectra for the same simulation parameters as in Fig.~\ref{fig02}; (a)-(c) electron energy spectra; (d)-(f): shell-resolved ion energy spectra (as indicated); (g)-(i): charge-state resolved ion energy spectra with and without recombination (as indicated) and corresponding ion charge distributions (insets).}
\end{minipage}
\end{figure}

The energy distributions of emitted electrons in Figs.~\ref{fig03}a-\ref{fig03}c show a transition from an exponential to a plateau-like structure.
In the VUV scenario (Fig.~\ref{fig03}a), only a small feature from multistep ionization below 4\,eV (see arrow) is observed within the otherwise smooth exponential spectrum, reflecting the dominant contribution from electron evaporation. For XUV excitation, multistep ionization produces a pronounced plateau with the characteristic sharp cutoff at the atomic photoline (at 22\,eV), see Fig.~\ref{fig03}b. Nevertheless, the spectrum still contains a sizable exponential component in the low energy range and beyond the atomic photoline. In contrast to that, the electron spectrum is completely dominated by the multistep plateau for the soft x-ray scenario in Fig.~\ref{fig03}c, where the evaporative part is negligible. It should be emphasized that the plateau contains a small step at about 60\,eV in this case. This feature results from the fact, that 3s and 3p ionization of the argon atoms produce separate multistep plateaus that are shifted by the difference of the corresponding ionization potentials (about 14\,eV) against each other. Such step in the spectrum is not possible for VUV excitation ($I_p^{\rm 3s}>\hbar\omega$) and less visible for the XUV case, where 3p ionization is much more dominant over the 3s channel as in the soft x-ray case.

The electron spectra are thus found to be very sensitive to the ionization processes and the excitation regime being probed and thus represent a useful sensor for the cluster expansion dynamics. Strongly exponential electron spectra indicate a strong contribution of hydrodynamic expansion while plateau structures typical for multistep ionization are a marker for Coulomb explosion.

We now turn to the analysis of the spectra of emitted ions, where the signatures are much more complex. The final ion energy spectrum for the VUV case (black dotted curve in Fig.~\ref{fig03}d) is weakly structured and exhibits a negative slope over almost the full energy range. This latter feature is typical for the hydrodynamic expansion regime~\cite{Murphy2008}. The shell-resolved analysis reveals that only ions from the outermost shell and a small portion of the second outer shell achieve high recoil energy whereas ions from inner shells acquire only very little energy. The weak modulation of the full spectrum results from different charge states of ions originating from the outermost shell, cf. Fig.~\ref{fig03}d in comparison with the charge-state resolved spectrum in Fig.~\ref{fig03}g. In addition, the charge-state resolved spectra for higher final ionization levels show isolated peak structures (green curve for $q$=2 and blue curve for $q$=3 in Fig.~\ref{fig03}g) which can be explained by electron-ion recombination. While highly charged ions from outer shells can escape quickly, the ionization levels of transiently produced high-z ions from inner shells are reduced by recombination (compare ion spectra before and after recombination in Fig.~\ref{fig03}g). Note that ions with a transient charge state $q=4$ are removed from the final spectra almost completely due to recombination. Strongly peaked and isolated features in charge-state resolved energy spectra thus indicate the presence of a dense nanoplasma (electrons are required to recombine) and are thus typical in the regime of hydrodynamic expansion. This picture is substantiated by recent experiments where such isolated features have been observed in charge-state resolved ion spectra from Xenon clusters in 62~nm pulses~\cite{Nagaya2010}.

For the XUV case, the total ion energy spectrum is much more structured and exhibits a strong oscillatory behavior. This structure can be traced back to different ion charge states originating mostly from the two outermost shells, see Fig.~\ref{fig03}e and \ref{fig03}h. The charge-state resolved contributions before recombination show nearly the same profiles (after rescaling the respective energy axis by the ion charge state). These features indicate that ions from outermost shells already expand similar to a regular Coulomb explosion. Ions from inner shells, on the other hand, still exhibit smooth energy spectra with negative slopes and much lower energies. The signal reduction at low energies for $q=2$ and $q=3$ results from recombination to the next lower charge states $(q \rightarrow q{\rm -1})$, which transfers the contributions to the energy spectra of charge state $q{\rm-1}$. Similarly, the signal gain at high energy in the spectra for $q=1$ and $q=2$ can be traced back to recombination from the next higher ionization stage $(q+1 \rightarrow q)$, leading to an additional step on the high energy side (see Fig.~\ref{fig03}h).  However, in the XUV scenario recombination is less efficient than for the VUV case, as the lower fraction of quasifree electrons and the higher temperature of the nanoplasma reduce the corresponding rates (compare spectra before and after recombination in Figs.~\ref{fig03}g and \ref{fig03}h). Nonetheless, the signatures from recombination reflect that a dense nanoplasma is still present in the XUV case, though hydrodynamic expansion and Coulomb explosion contribute with comparable strength.

The strongest oscillatory pattern in the ion energy spectra are found for soft x-ray excitation together with a clearly positive slope of the total energy spectrum at low energy, cf. Fig.~\ref{fig03}f. This feature reflects that Coulomb explosion is the dominant mechanism. The maxima of the shell-resolved distributions occur at nonzero energy values that gradually increase with shell number. Note that the peaks at higher ion energy (at about 160 and 260\,eV) contain contributions from the four outermost shells, representing a clear signature from overrun effects in the exploding cluster~\cite{LastJortner2001}. Inspection of the charge-state resolved spectra shows that recombination is negligible for this scenario. The latter two effects, however, can be resolved only with the full microscopic information available from the simulation.

From the above analysis it can be concluded, that also the ion spectra contain useful features to identify the type of the expansion and thus the ionization regime. The most important features that may be analyzed in corresponding experiments are the slope of the full ion energy spectrum at low energies and the presence of well isolated peaks in the energy spectra of high-z ions. The presence of a negative slope hints at hydrodynamic expansion effects, which are closely connected to extensive nanoplasma generation and ionization heating. A sizable contribution from Coulomb explosion, on the other hand, is indicated by a positive slope of the ion energy spectra at low energy, strong oscillatory pattern, and self-similar structures (after rescaling with charge state) of charge-state resolved ion energy spectra. These features are typical for the multistep ionization regime. Intermediate cases show step structures at the high energy side with respect to the maximum in the charge-state-resolved ion spectra that are produced from recombination of the respective higher charge states.
The identification of the strengths of hydrodynamic and Coulomb explosion effects from the ion spectra, however, is expected to be difficult in experiments, as focus averaging and a finite size distribution of the clusters induce additional blurring of the signatures.

\section{Summary and conclusions}
\label{sec:summary}

To summarize, we have investigated the transition from nanoplasma-driven hydrodynamic cluster expansion to mostly electrostatic Coulomb explosion for medium-sized Ar$_N$ in intense VUV, XUV, and soft x-ray pulses from free electron lasers. By analyzing the time-dependent cluster dynamics as well as final electron and ion spectra with quasi-classical simulations, we have shown that this transition is closely linked to a crossover in the ionization dynamics and electron emission processes. From this strong link the possibility arises to identify the dominant cluster expansion mechanism just from the evaluation of electron spectra, as exemplarily shown for three simulation scenarios:
While plateau-shaped structures arising from multistep ionization indicate Coulomb explosion, exponential electron spectra from thermal electron evaporation are a typical sign for hydrodynamic cluster expansion. The simultaneous measurement of electron and charge-state-resolved ion energy spectra thus offers a promising route towards a more detailed reconstruction of cluster expansion, overrun effects, and recombination processes from laser-cluster interactions in the short wavelength regime.

These results thus add a new flavor to FEL science with clusters. In previous work it has been pointed out that the characteristic ionization mechanisms and the absence of strong plasma heating in short-wavelength laser-cluster interactions can be exploited to measure the pulse fluence just from the width of the multistep plateau in the electron spectra~\cite{Moribayashi2009}. Here we have shown that even the expansion dynamics of nanosized target can be inferred from electron spectra. This would be of interest for applications that are closely related to the correlation between ionization and expansion dynamics of many-particle systems in intense FEL pulses, such as single-shot diffractive imaging or time resolved x-ray holography. Moreover, because of the conceptual similarities of key response mechanisms, cluster dynamics in intense short-wavelength FEL pulses can even be linked to the field of ultracold plasmas~\cite{Killian2010}.

\section{Acknowledgments}
Financial support by the Deutsche Forschungsgemeinschaft within the SFB 652/2 and computer time provided by the HLRN Computing Center are gratefully acknowledged.

\end{document}